\documentclass[12pt,preprint]{aastex}
\usepackage{epsfig}
\begin{document}
\shorttitle{Old HVS}
\shortauthors{Kollmeier et al.}
\newcommand{\msun}{M_{\odot}}
\newcommand{\kms}{\, {\rm km\, s}^{-1}}
\newcommand{\cm}{\, {\rm cm}}
\newcommand{\gm}{\, {\rm g}}
\newcommand{\erg}{\, {\rm erg}}
\newcommand{\kel}{\, {\rm K}}
\newcommand{\kpc}{\, {\rm kpc}}
\newcommand{\mpc}{\, {\rm Mpc}}
\newcommand{\seg}{\, {\rm s}}
\newcommand{\kev}{\, {\rm keV}}
\newcommand{\hz}{\, {\rm Hz}}
\newcommand{\etal}{et al.\ }
\newcommand{\yr}{\, {\rm yr}}
\newcommand{\mpyr}{{\rm mas}\, {\rm yr}^{-1}}

\newcommand{\gyr}{\, {\rm Gyr}}
\newcommand{\eq}{eq.\ }
\def\arcsec{''\hskip-3pt .}
 \shortauthors{Kollmeier et al.}
\shorttitle{Limits on Old HVS}

\received{}
\accepted{}

\title{SEGUE-2 Limits on Metal-Rich Old-Population Hypervelocity Stars In the Galactic Halo} 
\author{Juna A.\ Kollmeier\altaffilmark{1}, Andrew Gould\altaffilmark{2},  Constance Rockosi\altaffilmark{3}, Timothy C. Beers\altaffilmark{4}, Gillian Knapp\altaffilmark{5}, Jennifer A. Johnson\altaffilmark{2}, Heather Morrison\altaffilmark{6}, Paul Harding\altaffilmark{6}, Young Sun Lee\altaffilmark{4}, Benjamin A. Weaver\altaffilmark{7}, and the SEGUE-2 Collaboration}

\altaffiltext{1}{Observatories of the Carnegie Institution of Washington,
  813 Santa Barbara Street, Pasadena, CA 91101}
\altaffiltext{2}{The Ohio State University, 4055 McPherson Labs, Columbus, OH, 43210}
\altaffiltext{3}{Department of Astronomy and Astrophysics, University of California Santa Cruz, 201 Interdisciplinary Sciences Building (ISB)
Santa Cruz, CA 95064}
\altaffiltext{4}{Department of Physics \& Astronomy; CSCE:Center for the Study of Cosmic Evolution and JINA: Joint Institute for Nuclear Astrophysics, Michigan State University, E. Lansing, MI 48824 USA}
\altaffiltext{5}{Department of Astrophysical Sciences, Princeton University,
  Peyton Hall, Princeton, NJ 08544}
\altaffiltext{6}{Department of Astronomy,Case Western Reserve University, 10900 Euclid Ave, Cleveland, Ohio 44106}
\altaffiltext{7}{Center for Cosmology and Particle Physics, New York University, New York, NY 10003}
\begin{abstract}

We present new limits on the ejection of metal-rich old-population hypervelocity stars from the Galactic center (GC) as probed by the SEGUE-2 survey.  Our limits are a factor of 3-10 more stringent than previously reported, depending on stellar type.  Compared to the known population of B-star ejectees, there can be no more than 30 times more metal-rich old-population F/G stars ejected from the GC.  Because B stars comprise a tiny fraction of a normal stellar population, this places significant limits on a combination of the GC mass function and the ejection mechanism for hypervelocity stars.  In the presence of a normal GC mass function, our results require an ejection mechanism that is about 5.5 times more efficient at ejecting B-stars compared to low-mass F/G stars.

\end{abstract}

\section{Introduction}

Hypervelocity stars (HVS) have emerged as a promising way to probe the
dynamics and physical conditions at the Galactic center (GC).  Thus
far, the discoveries of HVS have predominantly been B-stars, with the
more recent addition of a small number of A-stars \citep{brown05,
  edelmann05, hirsch05, brown07a, brown07b,brown09}.  These objects
have been detected at large distances from the GC ($\sim 70$\,kpc); their flight times and spectral types are consistent with ages of less
than 100-$200\,$Myr.  Empirically, the known HVS constitute a relatively
young population.  While the youth of this population may reflect a
top-heavy mass function at the GC or an ejection mechanism that
greatly prefers stars of this mass range, it is also possible that the
increased difficulty of locating old-population \footnote{ We use the term ``old'' as a short-hand for ``long-lived''.  Similarly, ``young'' is used to denote ``short-lived'' stars.} HVS among a
predominantly old halo population has distorted our picture of the
conditions at the GC. In the absence of a complete kinematic census of
the Galactic halo, it is substantially more difficult to find
old-population HVS compared to young-population HVS.  This is because
the Galactic halo is composed of primarily old-population stars whose
red colors are very similar to those of low-mass HVS, which make photometric pre-selection for spectroscopy non-trivial
(Kollmeier \& Gould 2007).  By contrast, there are very few blue stars
in the halo, just a handful that are blue due to their youth (such as
``runaway B stars") and not many more old stars (like blue-horizontal
branch stars) that live in the halo ``legitimately".  Early type stars are more luminous and can be seen to larger
distances, and thereby probe larger volumes, relative to the more
typical G-stars that predominate in a normal mass function.  These factors
combine to make the photometric background of non-HVS to HVS
significantly reduced for early-type stars.  However, regardless of
the practical difficulties, it is essential to find the old-population
HVS (or determine that they do not exist) if one is to understand the
physics at the GC (Kollmeier \& Gould 2007).  The power of the
wide-field spectroscopic capability of the 2nd Sloan Extension for
Galactic Understanding and Exploration (SEGUE-2) can be directly
brought to bear on this problem.

In a previous paper, we used the entire available Sloan Digital Sky
Survey (SDSS) stellar database to place limits on the ejection of
metal-rich old-population hypervelocity stars (\citealt{kollmeier09}, hereafter K09).  Using turnoff stars
collected in SDSS throughout the lifetime of the survey, we analyzed
the high-velocity, $|V_G| >300 \kms$, distribution of stars.  The
underlying sample contained nearly 300,000 stars --- a motley
collection of calibration stars, failed quasar targets, and fiber-fill.  The sheer number of targets allowed us to obtain upper
limits to the total ejection rate of $\Gamma^{\rm F} < 60\, {\rm
  Myr^{-1}}$ and $\Gamma^{\rm G} < 300\, {\rm Myr^{-1}}$ for unbound F
and G stars, respectively.  Comparing to estimates of the B-star
ejection rate from the \citet{brown07b} survey, we obtained a relative
ejection rate of old to young metal-rich HVS consistent with a normal
stellar mass function.  To probe more deeply into this putative
population, we have implemented a well-defined target selection
algorithm within SEGUE-2 to more cleanly search for metal-rich
old-population HVS that have been ejected from the
GC.

In Section 2, we describe SDSS, SEGUE, SEGUE-2, and our target
selection.  In Section 3 we present the results of this selection and
derive our sensitivity to metal-rich old-population HVS from
which our new limits for the ejection of this population are derived.
In Section 4 we discuss these limits.

\section{Sample}
The sample we analyze here comprises a subset of objects targeted within the SEGUE-2 survey.
\subsection{SDSS, SEGUE and SEGUE-2}

SDSS-I was an imaging and spectroscopic survey that began routine
operations in April 2000, and continued through June 2005
\citep{fukugita96, gunn98, york00, hogg01, lupton01, smith02,
  stoughton02, ivesic04, gunn06, tucker06}.  The SDSS and its
extensions used a dedicated 2.5m telescope \citep{gunn06} located at
the Apache Point Observatory in New Mexico. The Sloan Extension for
Galactic Understanding and Exploration (SEGUE) is one of the three key projects in the recently completed first extension of the Sloan
Digital Sky Survey, known collectively as SDSS-II. The SEGUE program, which ran from July 2005 to July 2008, obtained $ugriz$ imaging of
approximately 3500 deg$^2$ of sky outside of the SDSS-I footprint, with special attention being given to scans of lower Galactic latitudes ($|b| < 35\degr$) in order to better probe the disk/halo interface of the Milky Way. SEGUE obtained approximately 240,000 medium-resolution spectra of Galactic stars, selected to explore the nature of stellar populations from 0.5 kpc to 100 kpc \citep{yanny09}. SDSS-III, which is presently underway, has already completed the sub-survey SEGUE-2, an extension primarily intended to obtain stellar spectra for distant stars that are likely to be members of the outer Galactic halo.

The first seven public data releases from SDSS \citep{abazajian03, abazajian04, abazajian05, abazajian09,adelman06, adelman07,adelman08} have produced over 350,000 stellar spectra (and their derived atmospheric parameters, where possible). More than 120,000 stellar spectra obtained during the course of SEGUE-2 will be distributed as part of the next public data release, DR8.

The SEGUE Stellar Parameter Pipeline processes the wavelength- and
flux-calibrated spectra generated by the standard SDSS spectroscopic
reduction pipeline \citep{stoughton02}, obtains equivalent widths
and/or line indices for more than 80 atomic and molecular absorption
lines, and estimates radial velocities, T$_{\rm eff}$, log g, and [Fe/H] through the application of a number of approaches \citep{allendeprieto08, lee08a, lee08b}.  The spectral resolution of the survey is R$\sim$2000 over wavelengths 385-920 nm, yielding typical velocity errors of 4-10 $\kms$.

\subsection{Target Selection}

While the target selection method advocated by \citet{kg07} is relatively
complete, it would require many fibers on each SEGUE-2 plate.  To
achieve greater economy of fibers, we adopted the following rather stringent selection for HVS candidates in SEGUE-2.  The new selection criteria make use of proper motion information as well as
photometric metallicity information to pre-select stars that have
enhanced likelihood of being GC ejectees.  This experiment, designed to
locate the ``metal-rich contaminant'' to the Galactic halo originally
envisioned by \citet{hills88}, selects stars to have a high probability
of being high-metallicity, and to have proper motions consistent with
high space velocities from the GC. We select stars satisfying the following criteria:
\begin{itemize}
\item{0) $|b|>30^{\circ}$}
\item{I) $17<g_0<20$}
\item{II) $\mu > 8\,{\rm mas\,yr^{-1}}$}
\item{III) $v_{\rm tot} > 400\,{\rm km\,s^{-1}}$}
\item{IV) either}
\begin{itemize}
\item{IVa) $\mu_\perp < 6\,{\rm mas\,yr^{-1}}$ or}
\item{IVb) $v_\perp > 400\,{\rm km\,s^{-1}}$}
\end{itemize}
\item{V) either}
\begin{itemize}
\item{Va) $(0.35< (g-r)_0\leq 0.40)\cap(0.375 < (u-g)_0-2.5(g-r)_0 < 0.525)$ or}
\item{Vb) $(0.40< (g-r)_0\leq 0.60)\cap(0.225 < (u-g)_0-2.5(g-r)_0 < 0.425)$}
\end{itemize}
\end{itemize}
where $\mu$ is the total proper motion as determined by SDSS and
USNO-B \citep{gk04,munn04}, $\mu_\perp$ is the component of proper
motion perpendicular to the direction from the GC to the star, $v_{\rm
  tot}$ is the total (3D) velocity (assuming that the star is on a
radial orbit), and $v_\perp = \mu d$ is the transverse velocity
assuming that the measured proper motion is correct (velocities quoted here are Galactocentric).  There are several other criteria that are applied to all SEGUE-2 targets involving proper motions that are designed to remove poorly measured objects due to errors in USNO-B.  These are described in greater detail below.  However, the crucial
point at this stage is that these criteria lead to a dramatic
increase in rejected targets for $|b|<30$, which is what motivates
Criterion (0).

\begin{figure*}
\plotone{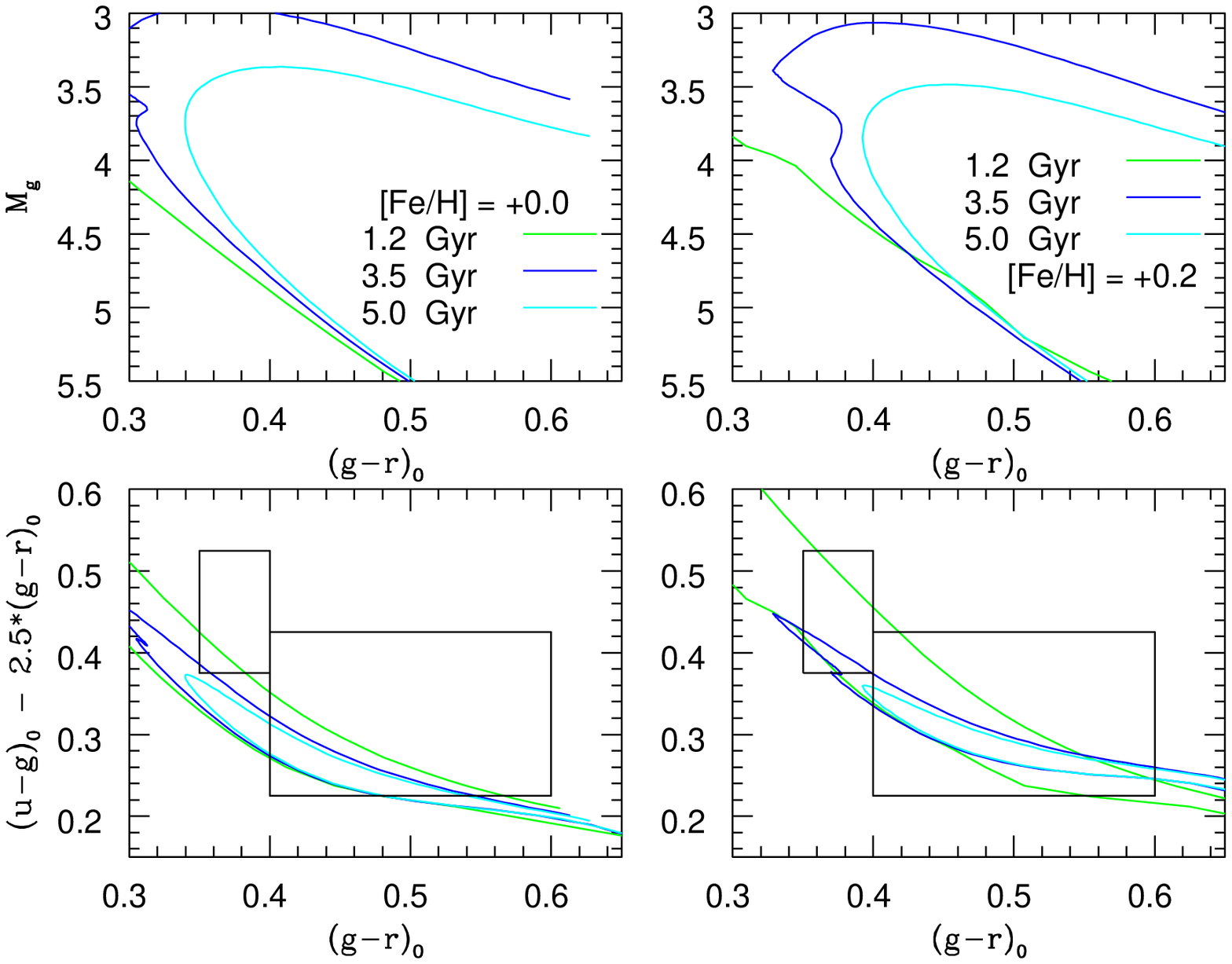}
\caption{\singlespace Color-magnitude and color-index diagram for theoretical isochrones computed by An et al. (2009) as a function of age and metallicity.  Left panels show results for a [Fe/H]=0.0 isochrone; right panels show a [Fe/H]=+0.2 isochrone.  Colors green, blue, and cyan show ages of isochrones corresponding to 1.2, 3.5, and 5.0 Gyr, respectively.  Our color selection boxes are superposed on the isochrones in the lower panels as black rectangles.  Our survey has sensitivity to solar and super-solar metallicity stars. }
\label{fig:isochrones} 
\end{figure*}

Criterion (I) restricts the sample to stars with modest $u$-band
photometric errors.  Criterion (II) eliminates many stars, while
preserving most HVS.  Given the $4\,{\rm mas\,yr^{-1}}$ errors,
criterion (III) eliminates a large fraction of stars surviving (II)
while only eliminating 7\% of HVS.  Criterion (IV) ensures a minimum
velocity in a direction consistent with an origin at the GC. Criterion
(V) is the most important.  It eliminates the overwhelming majority of
halo stars and selects for metal-rich stars.  The color index
$(u-g)_0-2.5(g-r)_0$ ``typically'' (i.e., at $g=19.4$) has errors of 0.105 mag,
which means that if the color-color relation (see
Fig~\ref{fig:isochrones}) is within a few hundredths of a magnitude of
the threshold, then roughly half of the stars will survive this cut.
Because of the relatively large error in this index, it is essential
to keep the boundary away from the bulk of the metal-poor population,
if one wants to avoid massive contamination.  Hence, the combination
of large errors and requirement of avoiding contamination dictates a
color-color cut that will eliminate about half the HVS.

We individually inspected a 20\% random sample of the HVS targets
rejected by the SEGUE-2 proper-motion selection algorithm on digitized POSS-I and POSS-II plates.  We found that essentially all of those rejected because they had less than 4 astrometric epochs were, in fact, spurious targets: they had nearby ($\la 5''$) neighbors that
were blended in POSS-I, thus corrupting the proper-motion
measurements.  Of the rejected targets that passed this test, almost
none had any recognizable problem.  These constitute 11\% of all
targets (excluding those rejected for too-few epochs).  In addition,
we estimate that $\la 2\%$ of HVS (i.e., random field stars) would
have a close neighbor that would lead to rejection as above.  Hence, we apply a 0.87 ($=1 - 0.11 -0.02$) correction factor to our
completeness, due to SEGUE-2 proper-motion selection.

Our color selection is motivated by the requirement that we use
minimal fibers per plate, so as not to interfere with any other program
(of which there are many; \citealt{yanny09}). To assess the quality of our
photometric pre-selection we rely on guidance from the theoretical
isochrones computed by \citet{an09}.  Figure~\ref{fig:isochrones}
shows several tracks of stars plotted in color-magnitude and
color-color space at several ages for a solar metallicity isochrone
and a super-solar isochrone.  As can be seen from the figure, our
selection recovers solar metallicity objects very well for colors $0.4
< (g-r)_0 < 0.5$.  At redder colors, the isochrones formally fall below
our selection, but by an amount that is smaller than the color-index
error.  At bluer colors, we retain substantial sensitivity only to
stars above the main-sequence turnoff. For higher-metallicity stars, the isochrones
shift upward in the color-index panels, improving our sensitivity at
all colors.  We are therefore sensitive to solar and super-solar metallicity ejectees.  These curves shift downward at lower
metallicity (not shown) away from our color-index selection region,
and we lose sensitivity to sub-solar stars, except above the turnoff.
This will be important to consider when computing our sensitivity to
ejectees in Section~\ref{sec:results}. At solar metallicity, the
isochrones straddle the color-index threshold in the region
$0.4<(g-r)_0<0.6$, deviating by an amount that is small compared to
the 0.105 color-index threshold over the entire range.  Hence, we
obtain roughly 50\% sensitivity to solar-metallicity G stars, at a
cost of only about 1 fiber per plate, thus not substantially
interfering with any other program.  And, at higher metallicity, the
recovered fraction is substantially higher.

\section{Results \label{sec:results}}
\subsection{Where are the Old-Population HVS?}
The selection described above resulted in a total of 361 target stars within 181 SEGUE-2 fields.  The velocity distribution of these targets is shown in Figure~\ref{fig:veldist}.  As can be seen from the velocity distribution, while there are high-velocity objects, none of the targets is moving in excess of, nor even close to, Galactic escape speed (e.g., Xue et al. 2007).  The selection therefore uncovered no metal-rich ejectees from this carefully constructed sample.  Our null result may reflect a true dearth of old-population HVS in the halo, or it may reflect a target selection algorithm that is too stringent.  In order to determine which is the case, we compute our sensitivity below.

\subsection{Sensitivity to Metal-Rich Ejectees}
We begin with a simple order-of-magnitude estimate of our survey sensitivity.  We first define what is meant by ``sensitivity''.  Consider a Galactocentric spherical shell of thickness $\Delta r$.  A star moving with speed $v$ out of the GC will spend a time $\Delta t = (1000 \kms/v) \times (\Delta r/ 1\,{\rm kpc})$ Myr traversing the shell.  The total sensitivity to objects at a given velocity is therefore the integral of this quantity over the survey magnitude range and volume probed including selection effects. For this initial example, consider targeting every star of a single stellar type, with absolute magnitude $M_g =4.5$, to the survey limiting magnitude of $g = 20$ over the full area of the SEGUE-2 survey. Such stars can be probed to distances of approximately $d=10^{0.2(g-M_g+5)} {\rm pc} = 12.5\,{\rm kpc}$ from the Sun.  Each SEGUE-2 spectroscopic plate has an area of $\Omega_{plate}=7\,{\rm deg}^2$, and there are a total of $N_{plates} = 181$.  The survey volume, as seen from the Sun, is therefore $V_{probed, \odot} = N_{plate}\Omega_{plate} d^3/3 = 265\, {\rm kpc}^3$.  This corresponds to a {\it Galactocentric} shell thickness of $\Delta r_{GC} = V_{probed, \odot} / 4 \pi d_g^2$, where $d_g$ is the corresponding Galactocentric distance of the survey limit.  At a fixed velocity, we can compute the time an ejectee spends in this shell, and therefore, the sensitivity as defined above.  For a fiducial survey direction in which $d_g$ corresponds to $10\, {\rm kpc}$, and a fiducial velocity of $500\kms$, the star spends a total of 0.42 Myr traversing the shell, and the survey would be sensitive to stars of this type if they were ejected at a rate of $\sim$ 2.3 Myr$^{-1}$.

\begin{figure*}
\plotone{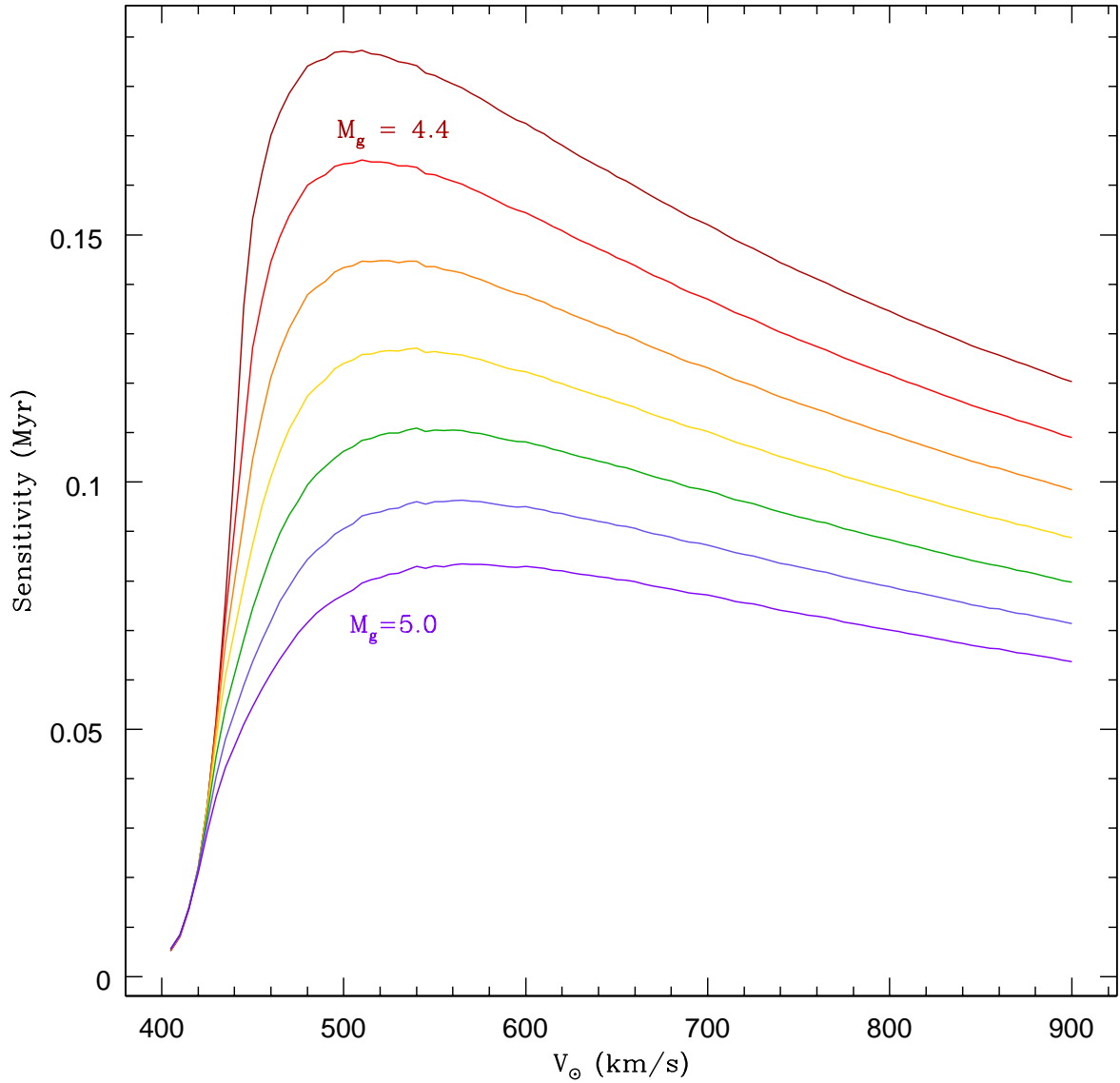}
\caption{\singlespace Sensitivity to HVS as a function of velocity at the solar circle and absolute magnitude of stellar population.  Curves from top to bottom show our sensitivity as a function of absolute magnitude ranging from $M_g=4.4$ (top) to $M_g=5.0$ (bottom) in increments of 0.1 magnitudes.}
\label{fig:sensorig} 
\end{figure*}

To compute the true survey sensitivity, we begin by fixing the
absolute magnitude at a range of values (as above), and also consider
a continuous range of ejection energies (parameterized by the HVS
velocity at the time it passes the solar circle).  For each of the
SEGUE-2 plates and for each apparent magnitude in our selection
interval, we compute $d$, $d_g$, and the local velocity, and then
simulate measurement of the star, allowing for proper-motion errors.
The resulting ``measurements'' are then fed into the same algorithm
that was used to select HVS candidates.  Summation over all fields and
apparent-magnitude intervals then yields the total sensitivity (in
Myr). We note the sharp contrast in character between our completeness
corrections and those made by \citet{kollmeier09}. That sample,
although huge in an absolute sense, contained only a small fraction of
potential HVS stars because it was not pre-selected.  Hence there was
a large completeness penalty. The present sample has stringent
selection, so that the only completeness factors are those due to
photometric errors (which scatter HVS in and out of our selection box)
and the SEGUE-2 proper-motion-error completeness factor of
0.87. Hence, our completeness factor (computed rigorously within our
code) is of order 50\%.  The result is shown in
Figure~\ref{fig:sensorig}.  At $M_g =4.5$ and $v_\odot = 500\,{\rm
  km\,s^{-1}}$, the sensitivity is 0.16 Myr, consistent (within a
factor of 3) with our order of magnitude calculation.

\begin{figure*}
\plotone{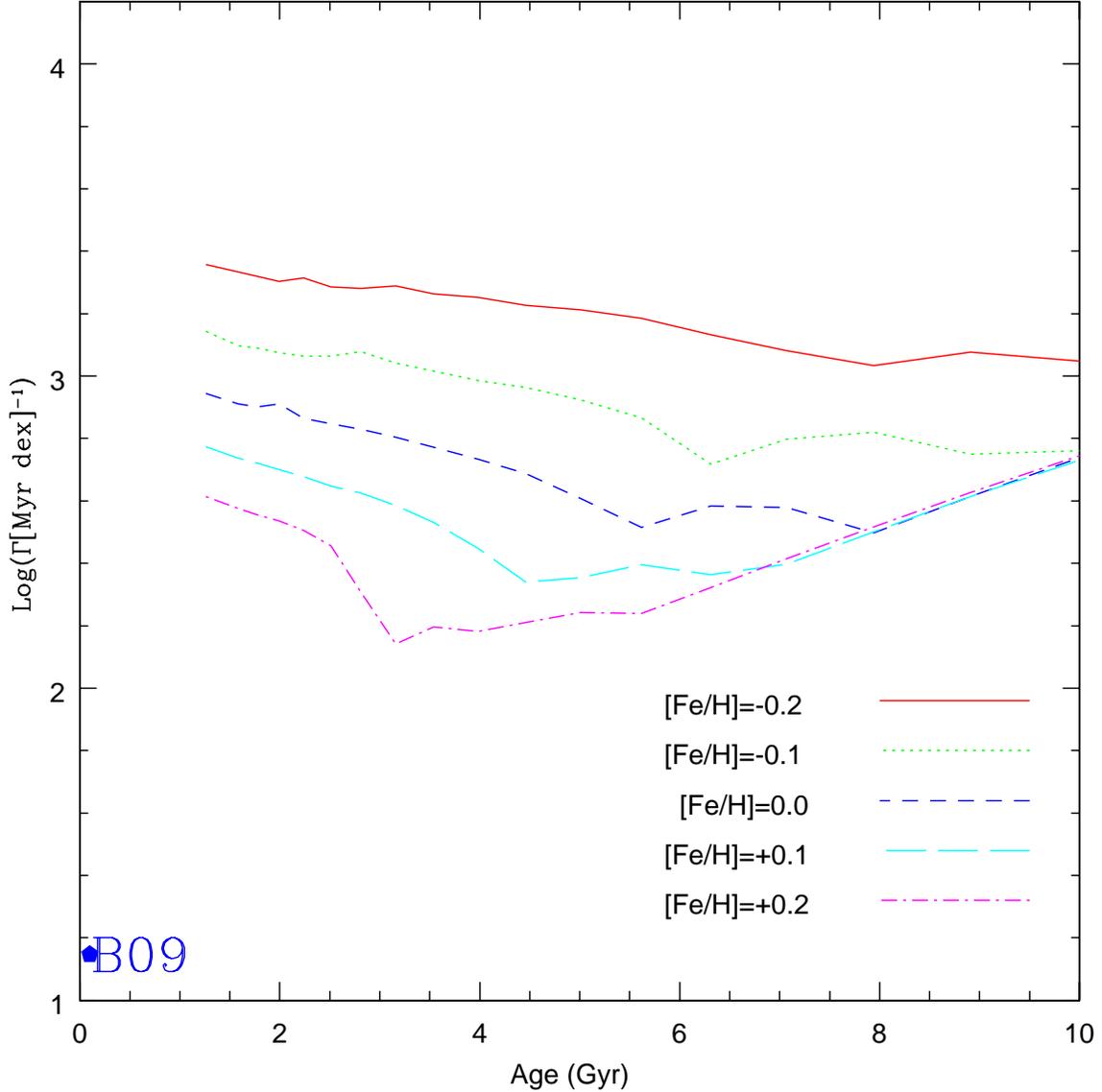}
\caption{\singlespace Limits on the ejection of old-population HVS as a function of metallicity and age, for velocities at the solar circle $v_\odot =
500\,{\rm km\,s^{-1}}$.  Solid, dotted, short-dashed, long-dashed, and dot-dashed lines correspond to metallicities of [Fe/H]$=+0.2, +0.1, 0.0, -0.1,$ and $-0.2$, respectively.   For comparison, we show the Brown et al. (2009) survey results, which probe a very different age regime compared to our survey.  Our ejection limits are most stringent for metal-rich stars; our sensitivity to sub-solar metallicity is significantly degraded.}
\label{fig:sensvage} 
\end{figure*}

Finally, to determine the sensitivity of the survey, we integrate the
functions shown in Figure~\ref{fig:sensorig} over isochrones
\citep{an09} of various ages and metallicities.  That is, at each mass
step (covering $\Delta\log m$), with specified $g-r$ color and $ugr$
color index, we multiply $\Delta\log m$ by the value indicated in
Figure~\ref{fig:sensorig} and by the fraction of stars that survive
our color selection (given the photometric errors), and sum over the
entire isochrone.  This yields a sensitivity $S$ (in Myr-dex).  We
therefore compute our sensitivity for isochrones of fixed age and a
range of mass. Since we detected no HVS in our survey, we have $1 -
e^{-n}n^0/n! \rightarrow 95\%$ confidence that there are fewer than
$n=3$ expected detections over the range of mass and age to which we
are sensitive.  That is, we obtain a lower limit $\Gamma > 3 S^{-1}$,
where $\Gamma$ is the rate of HVS ejections per Myr per dex.  In
Figure~\ref{fig:sensvage} we plot this lower limit on $\Gamma$ as a
function of population age for a range of metallicities at a fixed
velocity at the solar circle of $500\kms$.  As can be seen from the
figure, our strongest limits apply to super-solar metallicity stars with ages between 3-5 Gyr.  For comparison, we also show the rate derived
from the \citet{brown09} survey, which is sensitive to stars of ages
of roughly 150 Myr.  Rather than limits, that survey has secure
detections, so we plot it as a single point in the diagram.  However,
it is clear from this figure that our survey probes a very different
set of stellar parameters relative to Brown et al. (2009).

\subsection{Limits on Metal-Rich Ejectees}

\begin{figure*}
\plotone{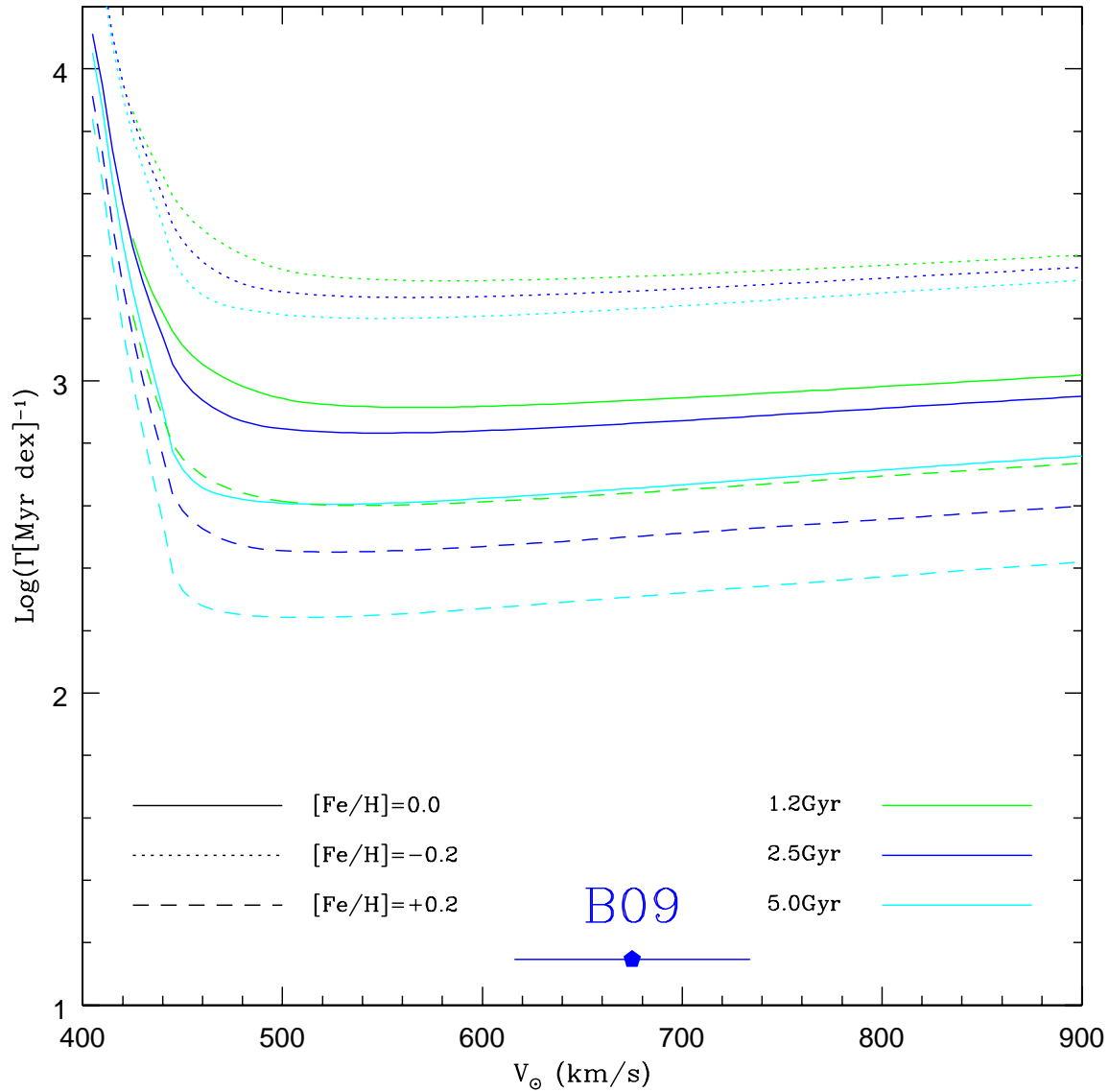}
\caption{\singlespace Limits on the ejection of old-population HVS as a function of metallicity, age, and velocity.  Solid, dotted, dashed lines correspond to metallicities of [Fe/H]=$0.0, -0.2$ and $+0.2$, respectively.  Colored lines show different population ages of green, blue, cyan to 1.2, 2.5 and 5.0 Gyr, respectively.  Each point of the solid curves is formed by integrating
over a set of corresponding points at the same velocity
in Figure~\ref{fig:sensorig}, weighting by the fraction that survive the color selection for each isochrone, and dividing 3 by the result. For comparison, we show the Brown et al. (2009) survey results.}
\label{fig:sens} 
\end{figure*}

Our survey uncovered no metal-rich hypervelocity stars.  As we have just demonstrated, our well-defined selection allows us to convert this null detection into an upper limit on the ejection rate as a function of stellar type. In Figure~\ref{fig:sens}, we show our ejection limits as a function of velocity at the solar circle for several population ages and metallicities.  We also show for comparison the results of the Brown et al. (2009) survey over the velocity range these HVS were observed.  At sub-solar metallicities, our limits are no more constraining than the K09 results (slightly less stringent in fact;  K09 found that there can be at most 100 times the ejection of old-population stars as young population stars and we find a limit of 120).  However, at solar metallicity and above, our limits are significantly stronger.  For these stars, there can be no more than a factor of 30 times more solar-metallicity, low-mass stars relative to B-star ejectees.  

At $\sim 500\kms$ our limit on the ejection of solar-metallicity stars from Figure~\ref{fig:sens} is roughly $\Gamma^{\rm 5Gyr,solar} < 410 {\rm (Myr-dex)}^{-1}$.  At [Fe/H]$=+0.2$, our limits are improved to $\Gamma^{\rm 5Gyr,super-solar} < 175 {\rm (Myr-dex)}^{-1}$.  

\subsection{Comparison to Young HVS}
The \citet{brown09} sample has 14 detected HVS objects over a total survey area $\Delta \Omega = 5000\,{\rm deg}^2 $, which corresponds to an ejection rate of roughly 1.75 Myr$^{-1}$ over a mass range of roughly 3-4$\msun$ (0.12 dex), or $\Gamma^{\rm 100Myr}=14\,{\rm (Myr-dex)}^{-1}$. For a Salpeter mass function with shape $dN/d\log M \propto M^{-1.35}$, we naively expect a factor of 5 more stars per dex at $1\msun$ compared to $3.5\msun$ available at the GC for ejection.  Of critical importance, however, is the relative lifetimes of these two populations.  Our sensitivity is maximized to stars of ages $5\,$Gyr.  The \citet{brown09} survey is targeted for stars with lifetimes of roughly 150Myr.  Therefore we must consider this additional factor of {\it 33 times} more stars available for ejection, were they in fact accumulating at the GC for $5\,$Gyr. Therefore, provided the mass function is roughly Salpeter at the GC and that stars were accumulating for at least 5$\,$Gyr, there should be $\sim$165 times more old-population stars (which our survey targets) relative to B-stars available for ejection.  Our limits place the ejection rate at no more than $\sim 30$ times more old-population ejectees relative to the known population of young ejectees. These stars are either not present at the GC, they have sub-solar metallicity, or the ejection mechanism quite strongly prefers high-mass stars at high velocity, by a factor of roughly 5.5.

\section{Discussion and Conclusions}

We have implemented a well-defined target selection algorithm within the SEGUE-2 survey to search for metal-rich old-population hypervelocity stars.  We targeted over 300 stars and found no hypervelocity stars.  This null result allows us to place new limits on the ejection of metal-rich old-population HVS from the GC as a function of mass, metallicity, velocity, and population age.  Our results imply that either the mass function at the GC is top-heavy, or the mechanism for ejection of HVS significantly favors massive stars at the level of 1:5.5, or that the stars being ejected from the GC are sub-solar metallicity.  

\begin{figure*}
\plotone{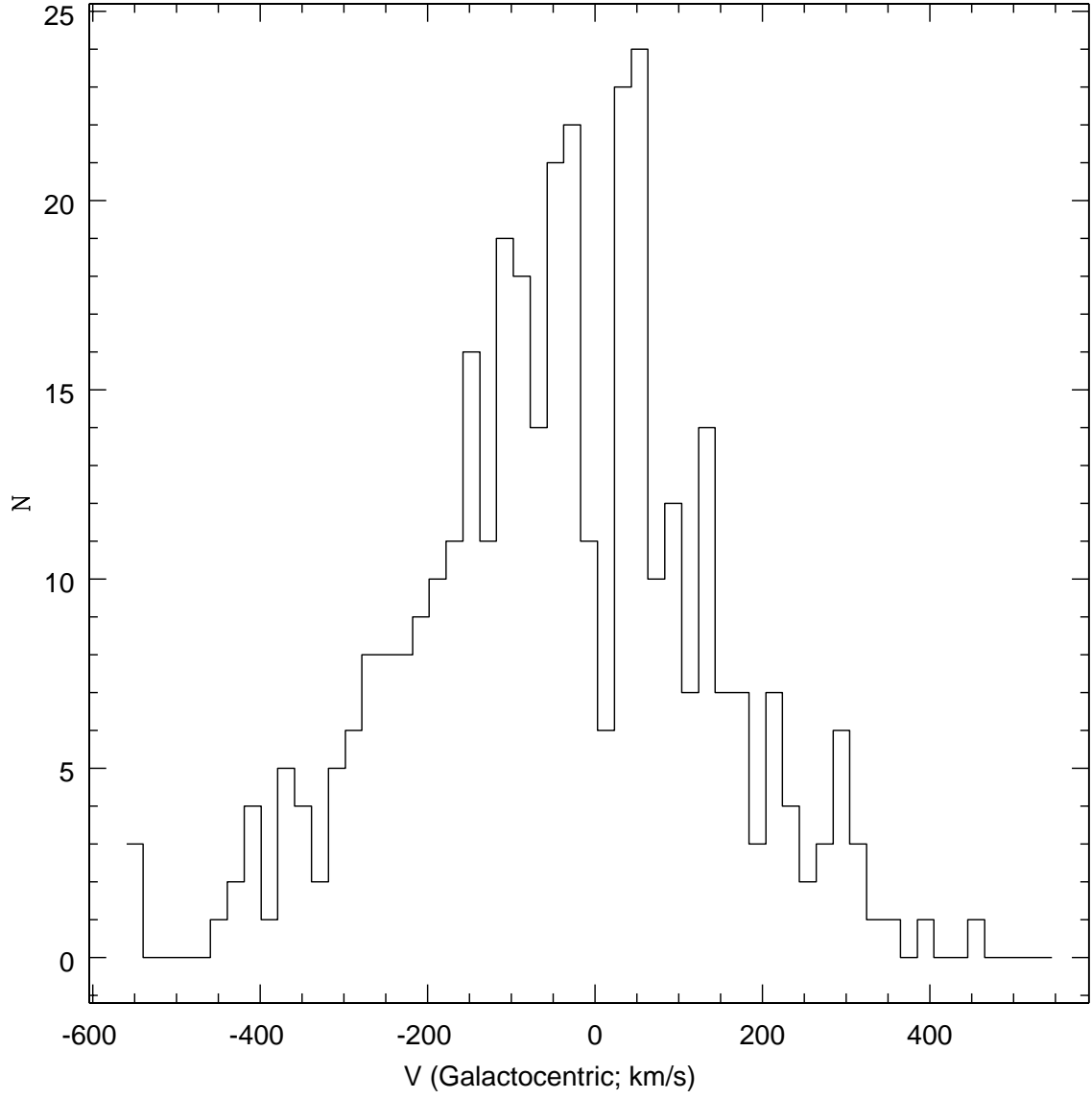}
\caption{\singlespace Galactocentric velocity distribution of HVS targets from SEGUE-2.  }
\label{fig:veldist} 
\end{figure*}

It is certainly plausible that the ejection mechanism favors high-mass stars, or more precisely, ejects low-mass stars at velocities too low to be detected by our survey.  It is possible that low-mass stars are being ejected at velocities below our survey threshold of $400\kms$, and some mechanisms, e.g., \citet{hills88}, would predict that low mass binaries would be ejected at lower velocity.  Indeed, detailed calculations by \citet{kenyon08} show that at $1\msun$, the predicted velocity distribution at $400\kms$ is strongly suppressed relative to more massive stars -- our result in Figure~\ref{fig:veldist} can be quantitatively compared to their predicted radial velocity distribution to determine whether the observations are consistent with the predictions for {\it bound} ejectees. Furthermore, high-mass stars are thought to have a binary fraction near unity \citep{pinstanek06}.  The binary fraction is indeed observed to be substantially less than unity at lower mass in the solar neighborhood \citep{dandm91, lada06}, and it is possible that conditions at the GC result in a similar F/G binary fraction.  Therefore, the classic binary disruption mechanism could very well favor high-mass stars in this scenario. Relaxing the assumption of a universal mass function, however, it is also possible that there are simply very few old stars at the GC.  For example, \citet{perets07} demonstrate that, for reasonable assumptions about the mass function and old-star binary-fraction, the ejection rate of 1$\msun$ stars should be approximately $5\times10^{-7} {\rm yr}^{-1}$ (but see \citealt{yandt03}), and could be substantially enhanced by the presence of a secondary massive perturber as envisioned by e.g., \citet{polnarev94}, \citet{baumgardt06}, and \citet{levin06} and not included in the \citet{kenyon08} calculations.

One mechanism that has been suggested is that the HVS have nothing at all to do with the black hole at the GC and the dynamics there, but rather they reflect debris of a satellite with pericenter close to the GC \citep{abadi09} and a large population of young stars.  The anisotropy of the known HVS potentially supports this interpretation.  A satellite galaxy would have the normal complement of old-population stars compared to young stars, and that would show up in our limits.  While we do not see these stars, if the satellite system had low metallicity similar to the current satellites of the Milky Way, it would not be detectable in our survey.  It is also possible that our SEGUE-2 pointings were not sufficiently aligned with the proposed direction of this debris.

The new limits derived here, in conjunction with the known HVS detections at higher mass, provide an important constraint on models of HVS ejection.  The target strategy implemented here represents an efficient and economical way to probe low mass populations ejected from the GC at modest spectroscopic follow-up cost.  Future spectroscopic surveys with similarly stringent target selection could strengthen these limits.  Deciphering the true distribution of GC ejectees as a function of mass, age, velocity, and position will continue to provide valuable insight into the still obscure Galactic center.

\acknowledgements{JAK acknoweledges the hospitality and support of the
  KITP during the program ``Building the Milky Way'' which was funded
  by NSF grant PHY-0551164.  Work by A.G.\ was supported in part by
  NSF grant AST-0757888.  Work by H.M.\ was supported in part by NSF
  grant AST-0098435.  T.C.B. and Y.S.L. acknowledge partial funding of this work from grants PHY 02-16783 and PHY 08-22648: Physics Frontier Center/Joint Institute for Nuclear
Astrophysics (JINA), awarded by the U.S. National Science Foundation.
Funding for SDSS-III has been provided by the
  Alfred P. Sloan Foundation, the Participating Institutions, the
  National Science Foundation, and the U.S. Department of Energy. The
  SDSS-III web site is http://www.sdss3.org/.  SDSS-III is managed by
  the Astrophysical Research Consortium for the Participating
  Institutions of the SDSS-III Collaboration including the University
  of Arizona, the Brazilian Participation Group, Brookhaven National
  Laboratory, University of Cambridge, University of Florida, the
  French Participation Group, the German Participation Group, the
  Instituto de Astrofisica de Canarias, the Michigan State/Notre
  Dame/JINA Participation Group, Johns Hopkins University, Lawrence
  Berkeley National Laboratory, Max Planck Institute for Astrophysics,
  New Mexico State University, New York University, the Ohio State
  University, University of Portsmouth, Princeton University,
  University of Tokyo, the University of Utah, Vanderbilt University,
  University of Virginia, University of Washington, and Yale
  University.

}

\end{document}